\documentclass[fleqn,usenatbib]{mnras}

\usepackage{newtxtext,newtxmath}

\usepackage[T1]{fontenc}

\DeclareRobustCommand{\VAN}[3]{#2}
\let\VANthebibliography\thebibliography
\def\thebibliography{\DeclareRobustCommand{\VAN}[3]{##3}\VANthebibliography}

\usepackage{graphicx}	
\usepackage{amsmath}	

\newcommand{\Msun}{\ensuremath{\mathrm{M}_\odot}}
\newcommand{\Zsun}{\ensuremath{\mathrm{Z}_\odot}}
\newcommand{\Lsun}{\ensuremath{\mathrm{L}_\odot}}
\newcommand{\Msunpyr}{\ensuremath{\Msun~\mathrm{yr^{-1}}}}
\newcommand{\kmps}{\ensuremath{\mathrm{km~s^{-1}}}}

\graphicspath{{./}{figures/}}

\title[Helium star mass loss and SN radio properties]{
Mass loss of massive helium star supernova progenitors shortly before explosion constrained by supernova radio properties
}

\author[T. J. Moriya and S.-C. Yoon]{
Takashi J. Moriya$^{1,2}$\thanks{E-mail: takashi.moriya@nao.ac.jp (TJM)}
and
Sung-Chul Yoon$^{3,4}$
\\
$^{1}$National Astronomical Observatory of Japan, National Institutes of Natural Sciences, 2-21-1 Osawa, Mitaka, Tokyo 181-8588, Japan \\
$^{2}$School of Physics and Astronomy, Faculty of Science, Monash University, Clayton, Victoria 3800, Australia \\
$^{3}$Department of Physics and Astronomy, Seoul National University, Seoul 08826, Republic of Korea \\
$^{4}$SNU Astronomy Research Center, Seoul National University, Seoul 08826, Republic of Korea
}

\date{Accepted 2022 May 3. Received 2022 April 30; in original form 2022 April 5}

\pubyear{2022}

\begin{document}
\label{firstpage}
\pagerange{\pageref{firstpage}--\pageref{lastpage}}
\maketitle

\begin{abstract}
Mass loss of massive helium stars is not well understood even though it plays an essential role in determining their remnant neutron-star or black-hole masses as well as ejecta mass of Type~Ibc supernovae. Radio emission from Type~Ibc supernovae is strongly affected by circumstellar matter properties formed by mass loss of their massive helium star progenitors. In this study, we estimate the rise time and peak luminosity distributions of Type~Ibc supernovae in radio based on a few massive helium star mass-loss prescriptions and compare them with the observed distribution to constrain the uncertain massive helium star mass-loss rates. We find that massive helium stars in the luminosity range expected for ordinary Type~Ibc supernova progenitors ($4.6\lesssim \log L/\Lsun \lesssim 5.2$) should generally have large mass-loss rates ($\gtrsim 10^{-6}~\Msunpyr$) in order to account for the observed rise time and peak luminosity distribution. Therefore, mass-loss prescriptions that predict significantly low mass-loss rates for helium stars in this luminosity range is inconsistent with the supernova radio observations. It is also possible that massive helium stars shortly before their explosion generally undergo mass-loss enhancement in a different way from the standard radiation-driven wind mechanism.
\end{abstract}




\section{Introduction}\label{sec:introduction}
Mass loss from massive stars plays a critical role in determining their fates. For example, black-hole mass distributions that can be measured by gravitational waves are strongly affected by assumed mass-loss rates of massive stars \citep[e.g.,][]{belczynski2010,higgins2021}. Type~Ibc supernovae (SNe~Ibc), which are core-collapse SNe without hydrogen features, somehow lose their entire hydrogen-rich envelopes before their explosion \citep[e.g.,][]{yoon2015}. Despite its critical importance, mass loss from massive stars remains to be one of the most uncertain processes in stellar evolution \citep[e.g.,][]{vink2021}.

Massive stars may explode as SNe at the end of their lives. SN explosions occur within the circumstellar matter (CSM) that is formed by the progenitors throughout their life time. Especially, radio emission from SNe is believed to be strongly affected by the CSM properties and radio observations of SNe have been used to constrain the mass-loss history of SN progenitors \citep[e.g.,][]{weiler2002}. For example, \citet[][]{moriya2021} recently used the rise time and peak luminosity distribution of SNe~II compiled by \citet[][]{bietenholz2021} to constrain the uncertain mass-loss prescriptions of red supergiants (RSGs).

Mass-loss rates of massive helium stars are as uncertain as those of RSGs and they are actively debated in recent years \citep[e.g.,][]{yoon2017,vink2017,sander2020b,sander2020}. When massive helium stars explode, they are observed as SNe~Ibc\footnote{
In this paper, we refer to hydrogen-poor massive stars as helium stars instead of Wolf-Rayet (WR) stars because we focus on those explode as SNe~Ibc. WR stars are characterized by strong emission lines from optically thick winds \citep[][]{crowther2007}. Not all SN~Ibc progenitors are expected to be WR stars because their winds would not be strong enough to be optically thick in many cases \citep[e.g.,][]{aguilera-dena2021}.
}. There have been several studies constraining the mass-loss rates of individual massive helium star SN progenitors through radio observations of individual SNe~Ibc \citep[e.g.,][]{berger2002,chevalier2006,soderberg2006,soderberg2010,wellons2012,maeda2013,milisavljevic2013,kamble2014,margutti2017,anderson2017,horesh2020,maeda2021}. In this work,
following the previous work by \citet[][]{moriya2021} for RSGs, we take the rise time and peak luminosity distribution of SNe~Ibc in radio compiled by \citet[][]{bietenholz2021} and constrain the general mass-loss properties of massive helium stars exploding as SNe~Ibc.

The rest of this paper is organized as follows. We first introduce the massive helium star mass-loss prescriptions based on which we estimate the expected radio properties of SNe~Ibc in Section~\ref{sec:masslossrates}. We briefly summarize the radio emission mechanism of SNe~Ibc in Section~\ref{sec:risepeak}. We compare the rise time and peak luminosity distributions of SNe~Ibc in radio and those expected from the massive helium star mass-loss prescriptions in Section~\ref{sec:comparison}. We discuss our results and conclude this paper in Section~\ref{sec:discussionandconclusions}.

\section{Massive helium star mass-loss rates}\label{sec:masslossrates}
We investigate three massive helium star mass-loss rate prescriptions at solar metallicity in this study. The first two prescriptions are from \citet{yoon2017}. \citet{yoon2017} provides massive helium star mass-loss rate prescriptions for WNE stars ($Y=1-\Zsun$, where $Y$ is the surface helium mass fraction) and for WC stars ($Y<0.9$). They are based on the observed mass-loss rates of WR stars \citep[][]{hamann2006,hainich2014,tramper2016}. The WNE mass-loss rate at solar metallicity is expressed as
\begin{equation}
    \dot{M}_\mathrm{WNE} = f_\mathrm{WR} 10^{-11.32} \left(\frac{L}{\Lsun}\right)^{1.18}~\Msunpyr, \label{eq:wne}
\end{equation}
where $f_\mathrm{WR}$ is the wind factor. The WC mass-loss rates at solar metallicity is
\begin{equation}
    \dot{M}_\mathrm{WC} = f_\mathrm{WR} 10^{-9.20}Y^{0.85} \left(\frac{L}{\Lsun}\right)^{0.83}~\Msunpyr. \label{eq:wc}
\end{equation}
The functional form of this WC mass-loss rate prescription is given by \citet[][]{tramper2016}, and \citet[][]{yoon2017} only multiplies the wind factor $f_\mathrm{WR}$. Because WC SN progenitor models typically have $Y\simeq0.2$ \citep{yoon2017}, we set $Y=0.2$ for the WC mass-loss rate throughout this paper. Following \citet{yoon2017}, we adopt $f_\mathrm{WR}=1.58$ which can explain the luminosity distribution of massive helium stars of different subtypes. Given that this prescription is based on the observed WR stars having $\log L/\Lsun > 5$  and $\log L/\Lsun \gtrsim 4.8$ for WNE and WC stars, respectively, applying this prescription to a less luminous helium stars is an extrapolation.

The last massive helium star mass-loss rate prescription we investigate is formulated by \citet{sander2020}. This mass-loss rate is based on theoretical stellar atmospheric calculations. The massive helium star mass-loss rate at solar metallicity is approximated as
\begin{equation}
    \dot{M}_\mathrm{Sander} = 10^{-4.06}\left(\log\frac{L}{L_0} \right)^{1.40}\left(\frac{L}{10L_0}\right)^{3/4}~\Msunpyr, \label{eq:sander}
\end{equation}
where $L_0=10^{5.06}\Lsun$. Note that this mass-loss prescription is obtained with a helium-rich surface composition resembling those of WNE stars, and cannot be directly applied to WC stars. However, \citet[][]{sander2020b} find in their model calculations that the mass-loss rate of WC stars is not much different from WNE stars.

\begin{figure}
	\includegraphics[width=\columnwidth]{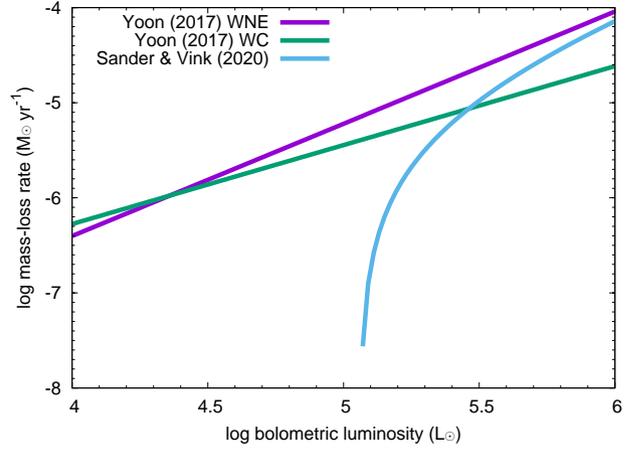}
    \caption{
    Mass-loss rate prescriptions of massive helium stars used in this study. They all assume solar metallicity.
    }
    \label{fig:masslossrate}
\end{figure}

Fig.~\ref{fig:masslossrate} illustrates the three mass-loss rate prescriptions used in this study. While the WNE and WC mass-loss rates from \citet{yoon2017} remains large at $\log L/\Lsun<5.0$, the massive helium star mass-loss rate of \citet{sander2020} drops at $\log L/\Lsun\simeq 5.5$ and insignificant mass loss is predicted at $\log L/\Lsun<5.0$.

The typical ejecta mass of SNe~Ibc which have massive helium progenitors is between $\simeq 1~\Msun$ and $\simeq 5~\Msun$ \citep[e.g.,][]{lyman2016,taddia2018}. If we assume a typical remnant mass of 1.4~\Msun, the massive helium star SN progenitor mass just before the explosion is between $\simeq 2.4~\Msun$ and $\simeq 6.4~\Msun$. The corresponding luminosity range is $4.6\lesssim \log L/\Lsun \lesssim 5.2$ \citep[][]{yoon2017}. Thus, we can probe the massive helium star mass loss in this luminosity range through SN radio properties. \citet{sander2020} predicts insignificant massive helium star mass loss in this luminosity range, while the WNE and WC mass-loss rates from \citet{yoon2017} suggest the massive helium star mass-loss rates larger than $10^{-6}~\Msunpyr$.

\begin{figure}
	\includegraphics[width=\columnwidth]{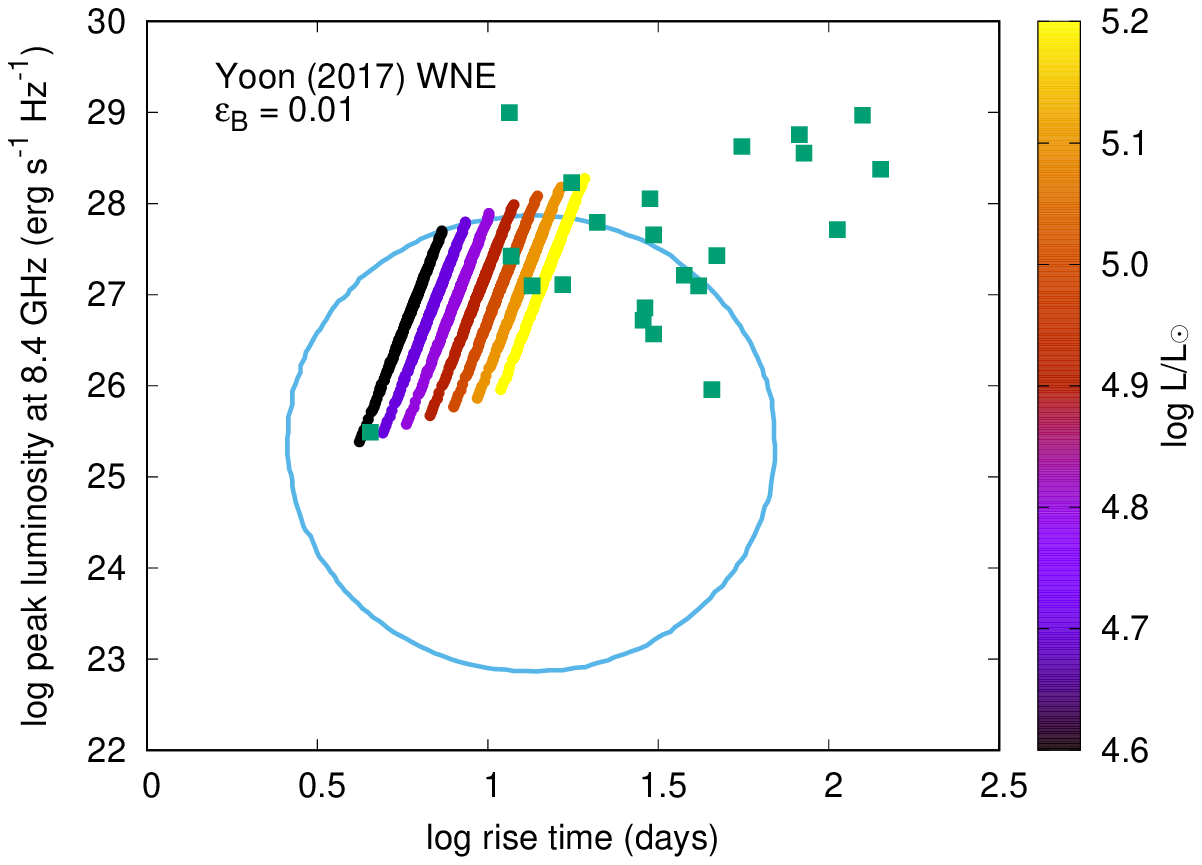}
	\includegraphics[width=\columnwidth]{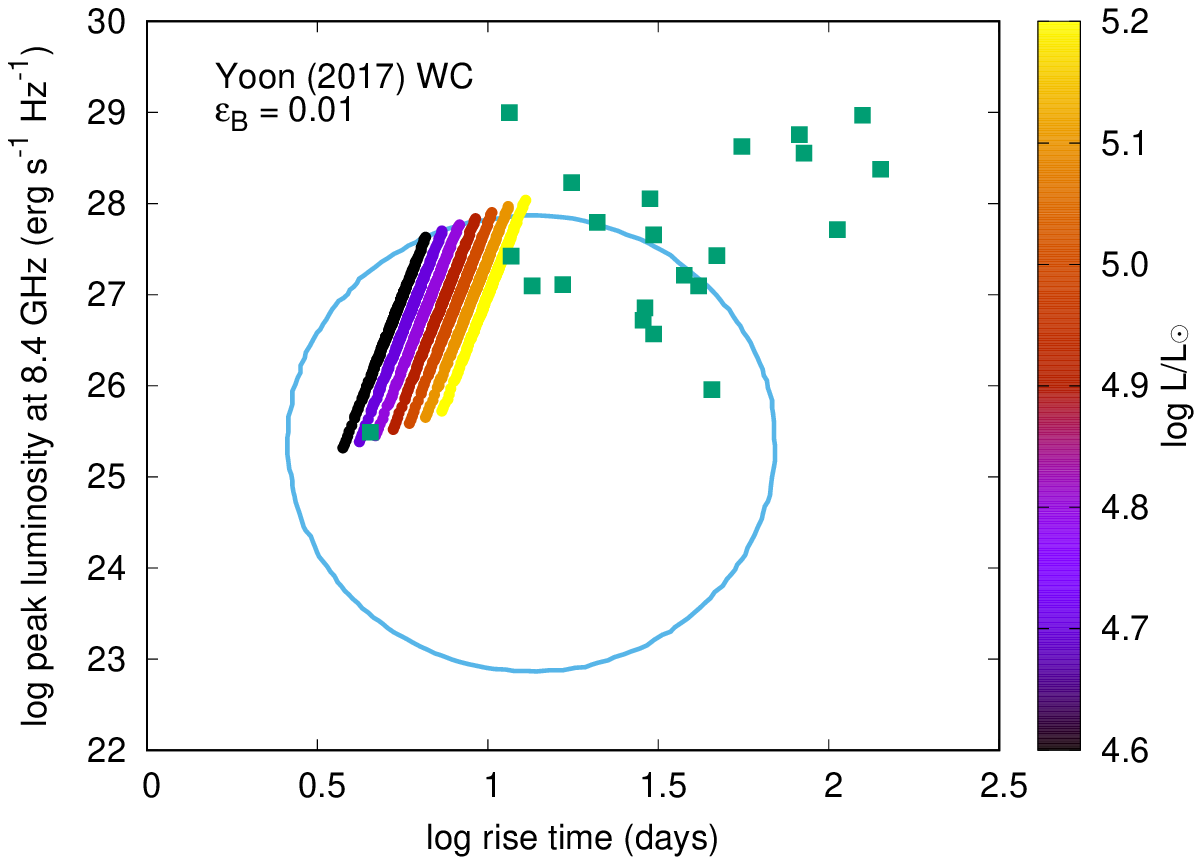}
	\includegraphics[width=\columnwidth]{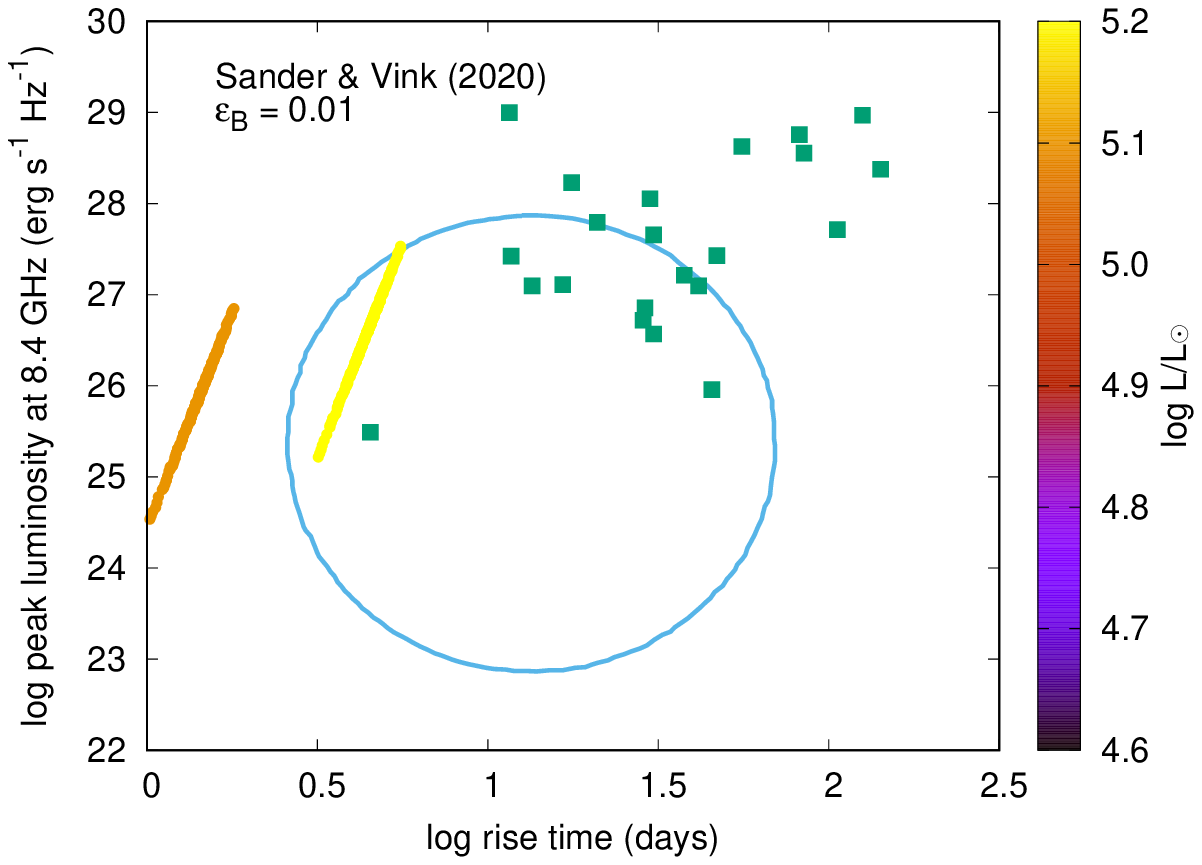}
    \caption{
    Rise time and peak luminosity of SNe~Ibc at 8.4~GHz. The green squares are the observed rise times and peak luminosities of SNe~Ibc presented by \citet{bietenholz2021}. The blue circle shows the region where the 68 per cent of SNe~Ibc are estimated to be located and it is obtained by taking all the observations including non detection into account \citep{bietenholz2021}. Each colored line shows an expected range of rise time and peak luminosity from a massive helium star of a given luminosity indicated by the color. The expected range is obtained by changing the ejecta mass ($0.5-5~\Msun$) and explosion energy ($0.5-10$~B). Although the ejecta mass can be determined by the progenitor luminosity in principle, we show all possibilities with $0.5-5~\Msun$ to account for uncertainty in stellar evolution theory. The top panel is for the WNE star mass-loss rate by \citet[][Eq.~\ref{eq:wne}]{yoon2017}, the middle panel is for the WC star mass-loss rate by \citet[][Eq.~\ref{eq:wc}]{yoon2017}, and the bottom panel is for the massive helium star mass-loss rate by \citet[][Eq.~\ref{eq:sander}]{sander2020}.
    }
    \label{fig:standard}
\end{figure}

\begin{figure}
	\includegraphics[width=0.86\columnwidth]{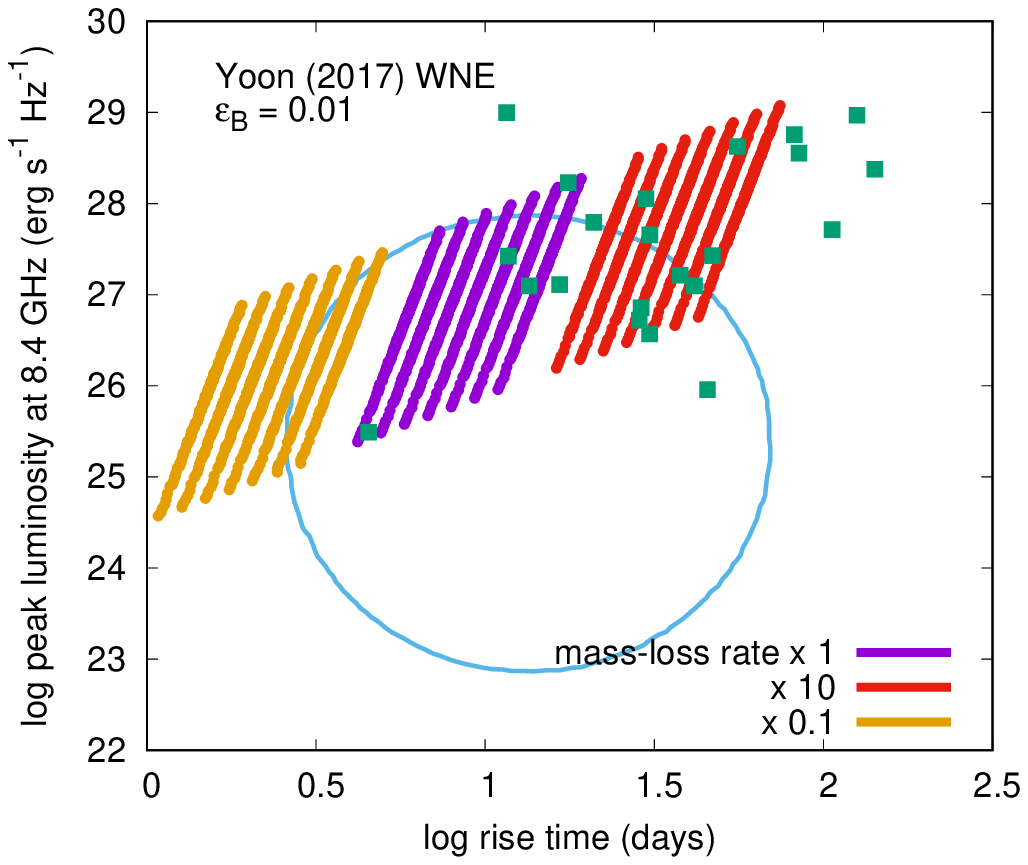}
	\includegraphics[width=0.86\columnwidth]{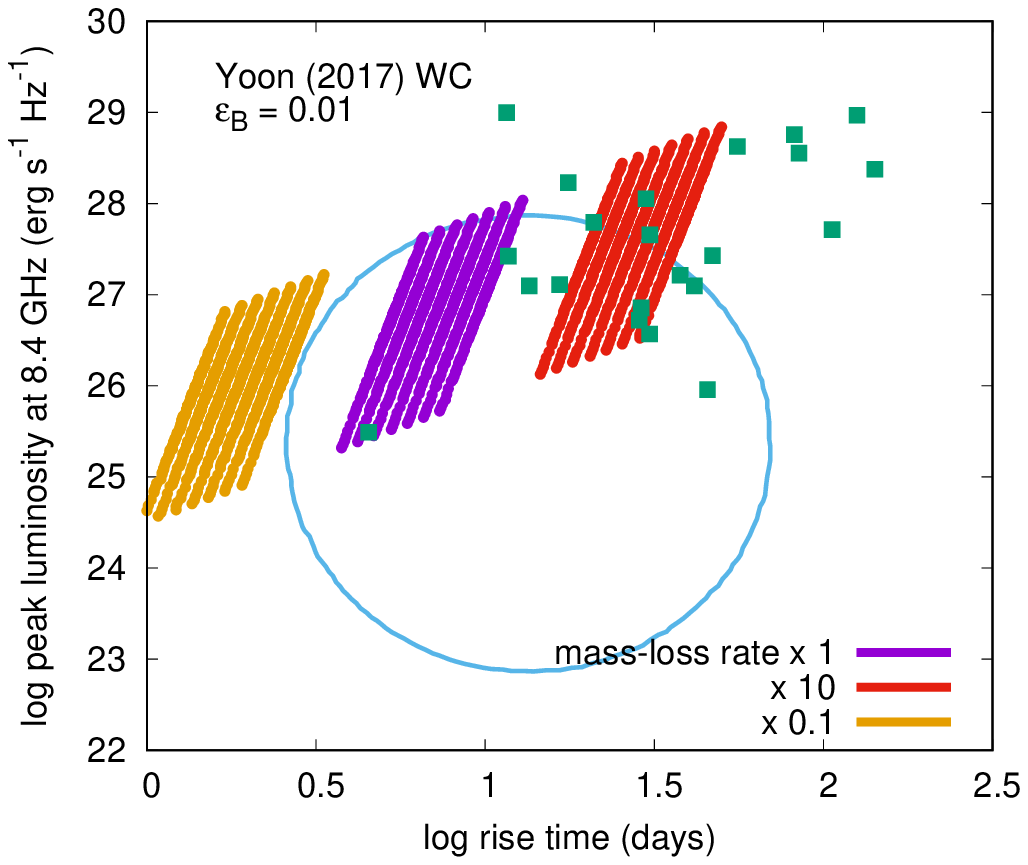}
	\includegraphics[width=0.86\columnwidth]{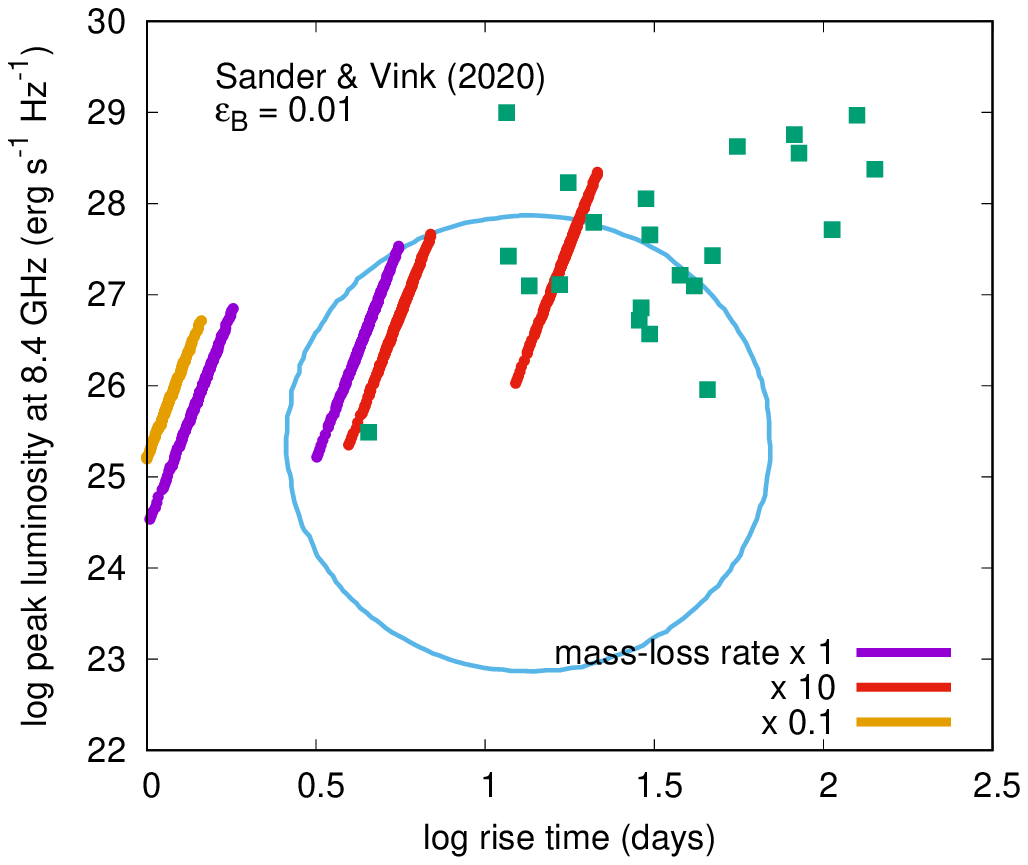}
    \caption{
    Same as Fig.~\ref{fig:standard}, but those obtained by changing the mass-loss rates by factors of 10 and 0.1.
    }
    \label{fig:factor10}
\end{figure}

\section{Rise time and peak luminosity of supernovae in radio}\label{sec:risepeak}
The synchrotron emission from electrons accelerated at the forward shock is assumed to be dominant in radio frequencies in SNe \citep[e.g.,][]{fransson1998,bjornsson2004}. We estimate the synchrotron luminosity in the same way as in our previous study of SNe~II and RSG mass loss \citep{moriya2021} and obtain the expected rise time and peak luminosity distribution from the massive helium star mass-loss prescriptions. We estimate the radio properties at 8.4~GHz in order to compare our estimates with the observations compiled at 8.4~GHz by \citet{bietenholz2021}.
The rise time and peak luminosity of SNe in radio are strongly affected by CSM density determined by mass loss of their progenitors. The rise time is determined by the timescale on which the optical depth of the synchrotron emission becomes unity. The peak luminosity increases as the CSM density increases. For a given SN ejecta properties, the rise time and peak luminosity of SNe in radio both become larger with higher CSM density, allowing us to constrain the CSM density and the progenitor mass-loss rates.
We refer to \citet{moriya2021} for the details of the radio luminosity formulation. Here, we only discuss important assumptions and differences between SNe~II and SNe~Ibc. 

Important parameters determining the SN radio properties are the fraction of kinetic energy used for electron acceleration ($\varepsilon_e$) and for magnetic field amplification ($\varepsilon_B$). Following the previous studies, we adopt $\varepsilon_e = 0.1$ and $\varepsilon_B = 0.01$ \citep[e.g.,][]{bjornsson2004,maeda2012,kamble2016}. The effects of $\varepsilon_e$ and $\varepsilon_B$ are discussed in \citet{moriya2021}.
Briefly, the fraction of $\varepsilon_e$ and $\varepsilon_B$ affects the rise time and peak luminosity of SN radio light curves. For a fixed $\varepsilon_e$, the rise time and peak luminosity increase as $\varepsilon_B$ increase.
However, the exact values of $\varepsilon_e$ and $\varepsilon_B$ do not affect our conclusions if they are changed in a reasonable range.
This is because we mainly discuss the overall distribution of the rise time and peak luminosity in radio. We have a large dispersion in the distribution that is caused by the diversity in ejecta properties and progenitor mass-loss rates. Because of the large dispersion, changing $\varepsilon_e$ and $\varepsilon_B$ in a reasonable range do not make a significant difference in the expected distribution compared with the observed distribution.

We consider synchrotron self-absorption (SSA) as the main physical process determining the rise time and peak luminosity of SNe~Ibc in radio \citep[][]{chevalier1998}. We ignore free-free absorption and inverse Compton cooling in this work. Free-free absorption is not likely to have significant effect in radio properties of SNe~Ibc because of their low CSM density and large forward shock velocity \citep[e.g.,][]{chevalier2006,maeda2012}. The inverse Compton cooling can reduce the peak radio luminosity by a factor of a few \citep{maeda2012}, but the small amount of the reduction does not affect our conclusion.

The radius and velocity of the forward shock are estimated by the self-similar solution formulated by \citet{chevalier1982}. We assume that the SN ejecta density structure ($\rho_\mathrm{ej}$) with two power-law components, i.e., $\rho_\mathrm{ej}\propto r^{-1}$ inside and $\rho_\mathrm{ej}\propto r^{-10}$ outside, which is suitable for stripped-envelope SNe \citep{matzner1999}. We assume the ejecta mass range of $0.5-5~\Msun$ and the explosion energy range of $0.5 - 10~\mathrm{B}$, where $1~\mathrm{B} = 10^{51}~\mathrm{erg}$, in order to cover the ejecta mass and explosion energy range of SNe~Ibc \citep[e.g.,][]{lyman2016,taddia2018}.

The CSM structure for a given mass-loss rate $\dot{M}$ is set as
\begin{equation}
    \rho_\mathrm{CSM} (r) = \frac{\dot{M}}{4\pi v_\mathrm{CSM}} r^{-2},
\end{equation}
where $v_\mathrm{CSM}$ is the CSM wind velocity. Because we investigate mass loss from massive helium stars in this work, we assume $v_\mathrm{CSM}=1000~\kmps$ which is a typical massive helium star wind velocity \citep[][]{hamann2006,hainich2014}.

\section{Massive helium star mass loss and supernova radio properties}\label{sec:comparison}
Fig.~\ref{fig:standard} shows the rise time and peak luminosity distributions at 8.4~GHz for the three massive helium star mass-loss prescriptions. The region expected from the WNE star mass-loss rate of \citet{yoon2017} covers the top part of the region estimated by the observations. The region expected from the WC star mass-loss rate is similar to that from the WNE star mass-loss rates, although the region covered by the WC star mass-loss rate is smaller. This is because the WC star mass-loss rate is less dependent on luminosity than the WNE star mass-loss rate (Fig.~\ref{fig:masslossrate}). The weaker luminosity dependence means smaller mass-loss rate diversity in the massive helium star SN progenitor luminosity range. Thus, it results in the smaller region covered in the rise time and peak luminosity.

While the rise time and peak luminosity expected from the WNE and WC star mass-loss rates of \citet{yoon2017} are within the range of those observed for SNe~Ibc, the rise time and peak luminosity estimated by the massive helium star mass-loss rate of \citet{sander2020} is mostly out of the expected region. With the \citet{sander2020} prescription, only the most luminous massive helium star SN progenitors with $\log L/\Lsun \simeq 5.1-5.2$ are expected to cover a small fraction of the observed distribution. This is simply because of the very small mass-loss rates expected for massive helium stars in the SN progenitor luminosity range of $\log L/\Lsun \simeq 4.6-5.2$ by the prescription of \citet{sander2020}. In addition, the mass of the brightest massive helium star SN progenitors with $\log L/\Lsun \simeq 5.1-5.2$ is $\gtrsim 5~\Msun$, and they are at the edge of the progenitor mass distribution of SNe~Ibc. The very small massive helium star mass-loss rates predicted by \citet{sander2020} do not explain the rise time and peak luminosity of most SNe~Ibc. We find that the SN~Ibc progenitor luminosity range needs to reach $\log L/\Lsun \sim 6$ to explain the most luminous SNe~Ibc in radio with the mass-loss rate of \citet{sander2020}. Thus, the SN~Ibc progenitor mass range needs to be from $\sim 5~\Msun$ to $\sim 30~\Msun$, which is much larger than observed.

Observationally, the massive helium star mass-loss rates for a given luminosity are scattered \citep[e.g.,][]{nugis2000,tramper2016} and the mass-loss rate formulae (Eqs.~\ref{eq:wne}, \ref{eq:wc}, and \ref{eq:sander}) only provide an average mass-loss rates for a given luminosity. In addition, the helium star mass-loss rate has uncertainty. \citet[][]{yoon2017} finds that  the luminosity distribution of WC stars can be well explained with $f_\mathrm{WR}$ = 1.58 in Equations~\ref{eq:wne} and~\ref{eq:wc}. However, this conclusion is based on the adopted dependence of the mass-loss rate on the WR luminosity and surface composition, which is poorly constrained. Note also that the adopted mass-loss rate prescription is based on the WR sample of the nearby universe, which is in the core helium-burning phase, and  might not be suitable for helium stars at the pre-SN phase. In order to take the scatter and uncertainty of the massive helium star mass-loss rate into account, we show the rise time and peak luminosity regions covered by changing each massive helium star mass-loss rate prescription within a factor of 10 in Fig.~\ref{fig:factor10}. We can find that the WNE and WC star mass-loss rates by \citet[][]{yoon2017} can cover broad ranges in the rise time and peak luminosity as observed in SN~Ibc with the major expected luminosity of massive helium star SN progenitors. However, the massive helium star mass-loss rates by \citet{sander2020} can only cover the observed range with the brightest massive helium star SN progenitors.

\section{Discussion and conclusions}\label{sec:discussionandconclusions}
We have shown that the WNE and WC mass-loss rate prescriptions by \citet{yoon2017} reasonably explain the broad rise time and peak luminosity range of SNe~Ibc in radio, while the prescription by \citet{sander2020} predicts too small mass-loss rates to explain the observation. However, our result does not mean that the prescription of \citet{sander2020} is not suitable for massive helium stars in general. The mass-loss history of massive helium stars that we can trace by SN radio observations is limited to shortly before the explosion because of the large massive helium star wind velocity (of the order of $1000~\kmps$) and forward shock velocity (of the order of $10000~\kmps$, e.g., \citealt{taddia2018}). During the typical radio rise times of SNe~Ibc ($10-100$~days), the forward shock reaches $\sim 10^{15}-10^{16}~\mathrm{cm}$. The CSM at $\sim 10^{15}-10^{16}~\mathrm{cm}$ is formed at $\sim 0.3-3~\mathrm{years}$ before the explosion. Thus, the mass-loss history we can probe in our study is that in a few years before the explosion\footnote{In the case of SNe~II as studied in \citet{moriya2021}, the mass-loss history we can trace is more than $\sim 30~\mathrm{years}$ before explosion because of slower RSG wind velocities ($\sim 10~\kmps$).}. Early SN~Ibc observations have revealed that massive helium star SN progenitors often experience mass-loss enhancement shortly before their explosions \citep[e.g.,][]{maeda2021}. The mass-loss enhancement forming a dense CSM is also suggested to be required to explain the luminous massive helium star SN progenitors that are directly identified (i.e., iPTF13bvn, SN~2017ein, and SN~2019yvr; e.g., \citealt{kilpatrick2021,jung2022}). Thus, it is still possible that massive helium stars have a very small mass-loss rate as predicted by \citet[][]{sander2020} until shortly before their explosion, and generally experience mass-loss enhancement from several years before explosions that can explain the rise time and peak luminosity distribution of SNe~Ibc in radio. Some possible mechanisms for such a mass-loss enhancement have been suggested \citep[][]{fuller2018}. However, note that with such a low mass-loss rate as predicted by \citet[][]{sander2020}, it would be very difficult to explain the relatively low ejecta masses of SNe~Ibc in terms of stellar evolution even if we consider binary interactions (see \citealt{yoon2017} for a related discussion).

The right-bottom part of the rise time and peak luminosity distribution in SNe~Ibc (blue circle in Figs.~\ref{fig:standard} and \ref{fig:factor10}) is not populated by the models presented in this work. The radio luminosity evolution with long rise times and low peak luminosity can be realized when free-free absorption is dominant \citep[e.g.,][]{moriya2021}. While the mass-loss rates we considered in this study (Fig.~\ref{fig:masslossrate}) are not high enough for the free-free absorption to be dominant, the mass-loss enhancement shortly before massive helium star explosions may sometimes make CSM dense enough for the free-free absorption to be dominant. SNe~Ibn with long rise time and low peak luminosity in radio can correspond to such a case. 

Radio SN observations provide valuable opportunities to investigate mass-loss history of SNe. Especially, they allow us to uncover mass-loss activities of massive helium stars shortly before explosion. This has important implications not only for stellar evolution theory but also for direct identifications of SN~Ibc progenitors, given that the mass-loss history immediately before the SN explosion can critically influence the optical brightness and color of SN~Ibc progenitors \citep[][]{jung2022}. Our radio SN observations are still biased to the bright end of the radio luminosity function \citep[][]{bietenholz2021}. The future radio facilities such as next-generation Very Large Array (ngVLA) will allow us to probe the broader range in the rise time and peak luminosity relation in SNe in radio. It will enable us to better uncover the mass-loss properties of SN progenitors shortly before their explosions.

\section*{Acknowledgements}
We thank the anonymous referee for constructive comments that improved this manuscript.
T.J.M. is supported by the Grants-in-Aid for Scientific Research of the Japan Society for the Promotion of Science (JP18K13585, JP20H00174, JP21K13966, JP21H04997).
S.-C.Y. is supported by the National Research Foundation of Korea (NRF) grant (NRF-2019R1A2C2010885).

\section*{Data Availability}
The data underlying this article will be shared on reasonable request to the corresponding author.
 


\bibliographystyle{mnras}
\bibliography{mnras} 






\bsp	
\label{lastpage}
\end{document}